\documentclass[reqno,12pt]{amsart}
 \usepackage{amssymb, latexsym, euscript}
 \newtheorem{theorem}{Theorem}[section]
 \newtheorem{proposition}[theorem]{Proposition}
 
 \newtheorem{lemma}[theorem]{Lemma}
 \newtheorem{definition}[theorem]{Definition}
 \newtheorem{corollary}[theorem]{Corollary}

\theoremstyle{definition}
\newtheorem{example}[theorem]{Example}
\newtheorem{remark}[theorem]{Remark}

\numberwithin{equation}{section}

\newcommand{\cC}{\mathcal{C}} \newcommand{\cP}{\mathcal{P}} \newcommand{\cT}{\mathcal{T}}    \newcommand{\cD}{\mathcal{D}}  \newcommand{\cR}{\mathcal{R}}   \newcommand{\sH}{\mathfrak{H}} \newcommand{\sL}{\mathfrak{L}}

\begin{document} \title[Towards theory of $\cC$-symmetries]{Towards theory of $\cC$-symmetries} \author[S.~Kuzhel]{S.~Kuzhel}
\author[V.~Sudilovskaya]{V.~Sudilovskaya}

\address{AGH University of Science and Technology \\ 30-059 Krak\'{o}w, Poland}
\email{kuzhel@agh.edu.pl}

\address{Kyiv Vocational College \\ Kyiv, Ukraina} \email{veronica.sudi@gmail.com}


\maketitle
\begin{abstract}
The concept of $\cC$-symmetry originally appeared in $\mathcal{PT}$-symmetric quantum mechanics is studied within the
Krein spaces framework.
\end{abstract}

\section{Introduction} 

Let $\mathfrak{H}$ be a Hilbert space with inner product $(\cdot,\cdot)$  linear in the first argument
and let $J$ be a non-trivial fundamental symmetry, i.e., $J=J^*$, $J^2=I$, and $J\not={\pm{I}}$.

The space $\mathfrak{H}$ endowed with the indefinite inner product
\begin{equation}\label{new1}
[f,g]=(J{f}, g)
\end{equation}
is called a Krein space $(\mathfrak{H}, [\cdot,\cdot])$.

A linear operator $H$  acting in $\mathfrak{H}$ is called $J$-\emph{self-adjoint} if ${J}H=H^*{J}$  and $J$-\emph{symmetric} if ${J}H{\subseteq}H^*{J}$.
 The condition of $J$-symmetricity of $H$
is equivalent to the relation \begin{equation}\label{ww30} [Hf,g]=[f, Hg], \qquad \forall{f,g}\in{D}(H). \end{equation}

During last two decades, the theory of $J$-self-adjoint operators has found successful applications in studying nonself-adjoint Hamiltonians of ${\mathcal{PT}}$-symmetric quantum mechanics (PTQM). The Hamiltonians of PTQM are not self-adjoint with respect to the initial Hilbert space's inner product $(\cdot,\cdot)$ but they possess a certain `more physical property' of symmetry, which does not depend on the choice of inner product (like $\mathcal{PT}$-symmetry property, see \cite{AK_Bender4, AK_Bender3} and references therein).
Typically, such nonself-adjoint operators can be interpreted as $J$-self-adjoint or $J$-symmetric operators for a suitable choice of Krein space. However, the self-adjointness of a Hamiltonian $H$ in a Krein space cannot be satisfactory because it does not guarantee the unitarity of the dynamics generated by $H$. In PTQM-related studies, this problem is solved by finding a new symmetry represented by a linear operator $\mathcal{C}$ and such that the new inner product $(\cdot,\cdot)_{\mathcal{C}}:=[\mathcal{C}\cdot,\cdot]$ ensures the self-adjointness
of $H$ in a (possible) new Hilbert space endowed with $(\cdot,\cdot)_{\mathcal{C}}$.

The description of a symmetry $\mathcal{C}$ for a given nonself-adjoint Hamiltonian $H$ is one of the key points in PTQM. Because of the complexity of the problem (since $\mathcal{C}$ depends on the choice of $H$), it is not surprising that the majority of the available formulae are approximative \cite{AK_Bender7, AK_Bender2} and the relevant mathematically study of the concept of $\cC$-symmetry is not yet developed well.

The concept of $\mathcal{C}$-symmetry in PTQM can be easily reformulated in the Krein space setting \cite{AK_AK}. The crucial point here is that the set of all operators $\cC$ in the Krein space $(\mathfrak{H}, [\cdot,\cdot])$ is in one-to-one correspondence with the set of all possible $J$-orthogonal pairs of maximal positive ${\mathfrak L}_+$ and maximal negative ${\mathfrak L}_-$ subspaces of $\mathfrak{H}$, where positivity (negativity) is understood with respect to the indefinite inner product $[\cdot, \cdot]$. The existence of a $\cC$-symmetry for a $J$-symmetric operator $H$ means that $H$ has a maximal dual pair $\{\sL_+, \sL_-\}$ of invariant subspaces.
This is an underlying mathematical structure that allows one to develop mathematically rigorous study of $\cC$-symmetries.

The present paper can be considered as an introduction to the theory of $\mathcal{C}$-symmetries, where the concept of operator $\mathcal{C}$ is consistently studied with the use of Krein spaces approach.

This paper is organized as follows. In section 2, after preliminary results related to the Krein space theory,
we discuss various definitions of operators $\cC$ which are typically used in literature. Further, the relation between operators $\cC$ and the concept of operator angles is studied and the difference between bounded and unbounded operators $\cC$ is explained.
Section 3 deals with the case of $J$-self-adjoint operators with $\cC$-symmetry. We show that $J$-self-adjoint operators with
non-zero resolvent set may only have a property of bounded $\cC$-symmetry, that is equivalent to the similarity of those operators to self-adjoint ones. In section 4, the method of construction of $\cC$ operators for $J$-symmetric operators with complete set
of eigenfunctions corresponding to real eigenvalues is presented. We characterize (possible) nonuniqueness of the corresponding unbounded operators $\cC$ with the use of extension theory of symmetric operators.

In what follows ${D}(H)$ and $\ker{H}$ denote, respectively, the domain and the kernel space of a linear operator $H$. The symbol $H\upharpoonright{\mathcal{D}}$ means the restriction of $H$ onto a set $\mathcal{D}$. Let $\sH$ be a complex Hilbert space.  Sometimes, it is useful to specify the inner product $(\cdot,\cdot)$ endowed with $\sH$. In that case the notation $(\mathfrak{H}, (\cdot,\cdot))$ will be used.

\section{Definition and general properties of operators $\cC$}

\subsection{Elements of the Krein space theory} Let $(\mathfrak{H}, [\cdot,\cdot])$ be a Krein space with a fundamental symmetry $J$. For the theory of Krein spaces and operators acting therein we refer the interested reader to \cite{AK_Azizov}.

A (closed) subspace $\mathfrak{L}$ of the Hilbert space $\mathfrak{H}$ is called {\it nonnegative, positive, uniformly positive} with respect to the indefinite innner product $[\cdot,\cdot]$ if, respectively,  $[f,f]\geq{0}$, \ $[f,f]>0$, \ $[f,f]\geq{\alpha}\|f\|^2, (\alpha>0)$ for all $f\in\mathfrak{L}\setminus\{0\}$. Nonpositive, negative and uniformly negative subspaces are introduced similarly.

In each of the above mentioned classes we can define maximal subspaces. For instance, a closed positive subspace $\mathfrak{L}$ is called \emph{maximal positive} if $\mathfrak{L}$ is not a proper subspace of a positive subspace in $\sH$. The concept of maximality for other classes is defined similarly.

Let ${\mathfrak L}_+$ be a maximal positive subspace.  Then its $J$-orthogonal complement (i.e., orthogonal complement with respect to the indefinite inner product $[\cdot, \cdot]$)
\begin{equation}\label{bebe91}
{\mathfrak L}_-={\mathfrak L}_+^{[\bot]}=\{f\in{\mathfrak H}\ :\ [f,g]=0,\ \forall{g}\in {\mathfrak L}_+\}
\end{equation}
is a maximal negative subspace of $\mathfrak{H}$, and the direct sum
\begin{equation}\label{e8}
{\cD}={\mathfrak L}_+[\dot{+}]{\mathfrak L}_-
\end{equation}
is a {\it dense set} in the Hilbert space $\mathfrak{H}$. The brackets in (\ref{e8}) indicates that
the subspaces ${\mathfrak L}_+$ and ${\mathfrak L}_-$  are $J$-orthogonal.

Denote \begin{equation}\label{AK9} \sH_+=\frac{1}{2}(I+J)\sH, \qquad \sH_-=\frac{1}{2}(I-J)\sH, \end{equation} where $J$ is the operator of fundamental symmetry which determines indefinite inner product in (\ref{new1}). It is easy to see that the $J$-orthogonal subspaces $\sH_{\pm}$ are maximal uniformly positive/negative with respect to $[\cdot,\cdot]$. Moreover, they are also orthogonal with respect to the initial inner product $(\cdot,\cdot)$  and
\begin{equation}\label{AK10}
\sH=\sH_+[\oplus]\sH_-.
\end{equation}
The decomposition (\ref{AK10}) is called \emph{the fundamental decomposition} of $\sH$.

The subspaces $\sL_\pm$  in (\ref{e8}) are  maximal positive/negative and their `deviation' from $\sH_{\pm}$  are described by a self-adjoint strong contraction\footnote{`strong contraction' means that $\|Tf\|<\|f\|, \ f\not=0$} $T$, which anticommutes with $J$. Precisely, \begin{equation} \label{e27} \sL_+=(I+T)\sH_+, \qquad   \sL_-=(I+T)\sH_-. \end{equation} The operator $T$ is called \emph{an operator of transition} from the fundamental decomposition (\ref{AK10}) to the direct sum (\ref{e8}) \cite{GKS}.

The collection of operators of transition admits a simple `external' description. Namely, a self-adjoint operator $T$ in $\sH$ is an operator of transition if and only if \begin{equation}\label{vvv1} \|Tf\|<\|f\| \quad (\forall{f}\in\sH, \ f\not=0), \qquad JT=-TJ. \end{equation}

The subspaces $J\sL_\pm$ are maximal positive/negative and theirs $J$-orthogonal direct sum
\begin{equation}\label{e8d}
J{\cD}=J{\mathfrak L}_+[\dot{+}]J{\mathfrak L}_-
\end{equation}
 is called \emph{dual} to (\ref{e8}). The transition operator corresponding to the dual decomposition (\ref{e8d})
 coincides with $-T$.

Denote by $P_{\sL_\pm} : \cD \to \sL_\pm$ the projection operators onto $\sL_\pm$ with respect to the decomposition (\ref{e8}).
The operators $P_{\sL_\pm}$ are defined on $\cD=\sL_+[\dot{+}]\sL_-$ and
$$
P_{\sL_\pm}f=P_{\sL_\pm}(f_{\sL_+}+f_{\sL_-})=f_{\sL_\pm}, \qquad f=f_{\sL_+}+f_{\sL_-}\in\cD, \ f_{\sL_\pm}\in\sL_{\pm}.
$$

It is known \cite{AK_AK} that
\begin{equation}\label{e30}
P_{\sL_+}=(I-T)^{-1}(P_{+}-TP_{-}), \qquad P_{\sL_-}=(I-T)^{-1}(P_{-}-TP_{+}),
\end{equation}
where $P_{+}=\frac{1}{2}(I+J)$ and $P_{-}=\frac{1}{2}(I-J)$ are orthogonal projection operators on $\sH_+$ and $\sH_-$, respectively.

\subsection{Definition and general properties of operators $\cC$.} Let a maximal positive subspace $\sL_+$ be given. Then its $J$-orthogonal complement ${\mathfrak L}_-={\mathfrak L}_+^{[\bot]}$ is a maximal negative subspace and we can consider the direct sum (\ref{e8}).

The operator $\cC$ associated with the direct sum (\ref{e8}) is defined as follows: according to (\ref{e8}), any element $f\in\cD$ admits the decomposition
\begin{equation}\label{new4}
f=f_{\sL_+}+f_{\sL_-}, \qquad f_{\sL_\pm}\in\sL_{\pm}.
\end{equation}
Then,
\begin{equation}\label{new5}
\cC{f}=\cC(f_{\sL_+}+f_{\sL_-})=f_{\sL_+}-f_{\sL_-}.
\end{equation}.

The operator $\cC$ associated with the direct sum (\ref{e8}) is \emph{$J$-self-adjoint and $J$-positive} \cite[Example 3.31]{AK_Azizov}. The latter means that $[\cC{f},f]>0$ for all nonzero $f\in{D(\cC)=\cD}$.

The set of operators $\cC$ in the Krein space $(\mathfrak{H}, [\cdot,\cdot])$ is in one-to-one correspondence with the set of all possible $J$-orthogonal direct sums (\ref{e8}). If $\cC$ is given, then the corresponding maximal positive/negative subspaces ${\mathfrak L}_\pm$ in (\ref{e8}) are recovered by the formula \begin{equation}\label{e9} \sL_\pm=\frac{1}{2}(I\pm\cC)\sH. \end{equation}

In view of (\ref{e30})  and (\ref{new5})
\begin{eqnarray}\label{bebe1}
\cC{f}=(P_{\sL_+}-P_{\sL_-})f=(I-T)^{-1}(P_{+}-TP_{-}-P_{-}+TP_{+})f= & & \nonumber  \\
(I-T)^{-1}(J+TJ)f=J(I+T)^{-1}(I-T)f & &
\end{eqnarray}
for all  $f\in{D}(\cC)$.

The definition of $\cC$ with the use of (\ref{new5}) is not always convenient because it requires a description
of the direct sum (\ref{e8}).  An additional analysis of (\ref{bebe1}) allows one to establish:

\begin{theorem}\label{ww1b} The following statements are equivalent: \begin{itemize}
  \item[(i)] an operator $\cC$ is determined by the direct sum (\ref{e8}) with the use
  of the formula (\ref{new5});
  \item[(ii)] an operator $\cC$ has the form
   $\cC=Je^Q$, where $Q$ is a self-adjoint operator in $(\sH, (\cdot,\cdot))$ such that $JQ=-QJ$;
  \item[(iii)] an operator $\cC$ satisfies the relation $\cC^2f=f$ for all $f\in{D(\cC)}$ and
  $J\cC$ is a positive self-adjoint operator in $\sH$.
\end{itemize}
\end{theorem}

The particular cases (the proof of equivalence of various relations between
the items (i)--(iii) of the theorem) have been considered in \cite{AK_AK}, \cite{AK_Bender}, and \cite{AK_Kuzhel}.

In what follows we will use assertions $(i) - (iii)$ of Theorem \ref{ww1b}
as \emph{(equivalent) definitions of operator} $\cC$.

This means that each operator $\cC$ has the form $\cC=Je^Q$ and,
simultaneously, the same operator $\cC$ can be determined by (\ref{new5}) for a
special choice of the direct sum (\ref{e8}).

Let us find the relationship between the operator of transition $T$ from the fundamental decomposition (\ref{AK10}) to
(\ref{e8}) and the self-adjoint operator $Q$.
In view of (\ref{bebe1}),
$$
e^Q=(I+T)^{-1}(I-T)
$$
or, after elementary transformations, $T=(I-e^Q)(I+e^Q)^{-1}$. Hence,
$$
T=-\frac{e^{{Q}/2}-e^{-{Q}/2}}{2}\left(\frac{
e^{{Q}/2}+e^{-{Q}/2}}{2}\right)^{-1}=-\frac{\sinh({Q}/2)}{\cosh({Q}/2)}=
-\tanh{\frac{{Q}}{2}}.
$$

\begin{corollary}\label{ww5}
The set of all operators $\cC$ is closed with respect to the calculation of the adjoint operator.
The operator $\cC^*$ is determined by the dual direct sum (\ref{e8d}).
\end{corollary}
\emph{Proof.}
In view of statement $(ii)$, the adjoint $\cC^*=[{J}e^Q]^*=e^Q{J}={J}e^{-Q}$ belongs to the set of operators $\cC$.
The operator $e^{-Q}$  corresponds to the transition operator $-T$ from the fundamental decomposition (\ref{AK10}) to
(\ref{e8d}). Therefore, $\cC^*$ is determined by the dual direct sum (\ref{e8d}).
\rule{2mm}{2mm}

\subsection{Relation between operators $\cC$ and operator angles.}

Recall that the operator angle $\Theta(\mathfrak{M}, \mathfrak{N})$ between subspaces  $\mathfrak{M}$ and $\mathfrak{N}$ measured relative to  $\mathfrak{M}$ is given by (see, e.g., \cite{AMT})
$$
\Theta(\mathfrak{M}, \mathfrak{N})=\arcsin\sqrt{I_{\mathfrak{M}}-P_{\mathfrak{M}}P_{\mathfrak{N}}}\upharpoonright_\mathfrak{M},
$$
where $I_{\mathfrak{M}}$ denotes the identity operator on ${\mathfrak{M}}$, and $P_{\mathfrak{M}}$ and $P_{\mathfrak{N}}$
stand for the orthogonal projections in $\sH$ onto $\mathfrak{M}$ and $\mathfrak{N}$, respectively.

By definition, the operator angle $\Theta(\mathfrak{M}, \mathfrak{N})$ is a bounded
non-negative operator on $\mathfrak{M}$ and
$$
0\leq\Theta(\mathfrak{M}, \mathfrak{N})\leq{\frac{\pi}{2}I}.
$$
This inequality can be specified for the cases where $\mathfrak{M}=\sH_\pm$ and $\mathfrak{N}=\sL_\pm$.
\begin{lemma}\label{bebe92}
Let $\sH_{\pm}$ and $\sL_\pm$ be determined by (\ref{AK9}) and (\ref{e27}), respectively. Then:
$$
\Theta(\sH_\pm, \sL_\pm)=\arcsin\sqrt{T^2(I+T^2)^{-1}}=\arcsin|T|\sqrt{(I+|T|^2)^{-1}},
$$
where $|T|=\sqrt{T^2}$ is the modulus of $T$.
\end{lemma}
\emph{Proof.} Let ${\mathfrak{M}}=\sH_+$ and $\mathfrak{N}=\sL_+$. Denote by $\widehat{P}_{\sL_+}$
the orthogonal projection in $\sH$ onto $\sL_+$. The orthogonal  projection $P_+$ in $\sH$ onto $\sH_+$
has the form $P_+=\frac{1}{2}(I+J)$.

By virtue of (\ref{bebe91}), $\sH=\sL_+\oplus{J}\sL_-$. Hence, any $f_+\in\sH_+$ can be decomposed:
$$
f_+=(I+T)y_++J(I+T)z_-=(I+T)y_+-(I-T)z_-, \quad y_+\in\sH_+, \ z_-\in\sH_-.
$$
Separation of the elements belonging to $\sH_\pm$, gives the system of equations
$y_+-Tz_-=f_+, \ T_+y_++z_-=0$, which enables one to establish that
$(I+T^2)y_+=f_+$ and $y_+=(I+T^2)^{-1}f_+$.
Using the decomposition of $f_+$, we obtain
$$
P_+\widehat{P}_{\sL_+}f_+=P_+(I+T)y_+=y_+=(I+T^2)^{-1}f_+.
$$
Therefore,
$$
(I-P_{+}\widehat{P}_{\sL_+})f_+=f_+-(I+T^2)^{-1}f_+=T^2(I+T^2)^{-1}f_+
$$
that completes the proof for $\Theta(\sH_+, \sL_+)$. The case $\Theta(\sH_-, \sL_-)$
is considered similarly.
\rule{2mm}{2mm}
\begin{corollary}\label{bebe96}
The operator angles $\Theta(\sH_\pm, \sL_\pm)$  satisfy the inequality
\begin{equation}\label{bebe93}
0\leq\Theta(\sH_\pm, \sL_\pm)\leq{\frac{\pi}{4}}I.
\end{equation}

The operator $\cC$ corresponding to the direct sum (\ref{e8}) of the subspaces $\sL_{\pm}$ is bounded
if and only if
$$
\|\Theta(\sH_\pm, \sL_\pm)\|=\max\{\lambda \ : \ \lambda\in\sigma(\Theta(\sH_\pm, \sL_\pm))\}<{\frac{\pi}{4}}.
$$
\end{corollary}
\emph{Proof.}
In view of Lemma \ref{bebe92}, $\Theta(\sH_\pm, \sL_\pm)$ are the functions of a nonnegative self-adjoint contractions
$|T|_{\pm}=|T|\upharpoonright_{{\sH_{\pm}}}$ acting in $\sH_{\pm}$. Precisely,
\begin{equation}\label{bebe94}
\Theta(\sH_\pm, \sL_\pm)=g(|T|_{\pm}), \qquad g(\lambda)=\arcsin\left(\frac{\lambda}{\sqrt{1+\lambda^2}}\right).
\end{equation}
This formula immediately leads to (\ref{bebe93}) since $0\leq{g(\lambda)}\leq\frac{\pi}{4}$ for  $\lambda\in[0,1]\supset\sigma(|T|_{\pm})$.
The second assertion of corollary \ref{bebe96} also follows from (\ref{bebe94}) and the fact that
an operator $\cC$ is bounded if and only if the corresponding operator of transition $T$
is an uniformly strong contraction (i.e., $\|T\|<1$).
\rule{2mm}{2mm}

An elementary calculation shows (see, e.g., \cite{KKM}) that
$$
|T|_{\pm}=\tan\Theta(\sH_\pm, \sL_\pm).
$$
On the other hand,
$$
|T|=|\tanh\frac{Q}{2}|=|T|_{+}P_++|T|_{-}P_-.
$$
Therefore, the relationship between the self-adjoint operator $Q$ in the formula $\cC=Je^Q$ and the operator angles
$\Theta(\sH_\pm, \sL_\pm)$ where $\sL_{\pm}$ are defined by $\cC$ is:
$$
|\tanh\frac{Q}{2}|=\tan\Theta(\sH_+, \sL_+)P_++\tan\Theta(\sH_-, \sL_-)P_-.
$$

\subsection{Bounded and unbounded operators $\cC$.}
There is an essential difference between the properties of bounded and unbounded operators $\cC$.
In particular, the spectrum of a bounded operator $\cC$ coincides with eigenvalues $\{-1, 1\}$, while the spectrum of unbounded $\cC$ additionally has the continuous part on $\mathbb{C}\setminus\{-1, 1\}$ \cite{AK_AK}.

If $\cC$ is \underline{bounded}, then: $\cC=Je^Q$, where $Q$ is a bounded self-adjoint operator in $\sH$; the subspaces $\sL_\pm$ in (\ref{e8}) turns out to be \emph{maximal uniformly} positive/negative in the Krein space $(\mathfrak{H}, [\cdot,\cdot])$; the direct sum (\ref{e8}) gives the decomposition of the whole space $\sH$:
\begin{equation}\label{e8b}
\sH={\mathfrak L}_+[\dot{+}]{\mathfrak L}_-;
\end{equation}
the given indefinite inner product $[\cdot, \cdot]$ generates infinitely many equivalent
 inner products $(\cdot,\cdot)_{\cC}$ in the Hilbert space $(\sH, (\cdot,\cdot))$,
  which are determined by the choice of bounded operators $\cC$:
 \begin{equation}\label{new7}
 (\cdot,\cdot)_{\cC}=[\cC\cdot, \cdot]=(J\cC\cdot, \cdot)=(e^Q\cdot, \cdot);
\end{equation}
 the subspaces ${\mathfrak L}_\pm$ in (\ref{e8b}) are mutually orthogonal with respect to $(\cdot,\cdot)_{\cC}$ and  $\cC$ turns out to be the fundamental symmetry operator in the Hilbert space $(\sH, (\cdot,\cdot)_{\cC})$.

\medskip

If $\cC$ is \underline{unbounded}, then: $\cC=Je^Q$, where $Q$ is an unbounded self-adjoint operator in $\sH$;
the subspaces $\sL_\pm$ in (\ref{e8}) are \emph{maximal} positive/negative in the Krein space $(\mathfrak{H}, [\cdot,\cdot])$; the direct sum (\ref{e8}) is a dense subset of $\sH$; the formula (\ref{new7}) determines
 infinitely many inner products $(\cdot,\cdot)_{\cC}$
defined on various sets $\cD=D(\cC)$; the linear spaces $\cD$ endowed with inner products $(\cdot,\cdot)_{\cC}$ are pre-Hilbert spaces;
the completion $\sH_{\cC}$ of $\cD$ with respect to $(\cdot,\cdot)_{\cC}$ does not coincide with $\sH$;
the Hilbert space $(\sH_{\cC}, (\cdot,\cdot)_{\cC})$ has the decomposition
\begin{equation}\label{e8c}
\sH_{\cC}=\hat\sL_+\oplus_{\cC}\hat\sL_-,
\end{equation}
where the mutually orthogonal\footnote{with respect to the inner product $(\cdot,\cdot)_{\cC}$}
subspaces $\hat\sL_\pm$ are the completion of $\sL_\pm$ with respect to $(\cdot,\cdot)_{\cC}$; the decomposition (\ref{e8c}) gives rise to the new Krein space $(\sH_{\cC}, [\cdot, \cdot]')$ with the indefinite inner product $$ [f, g]'=(f_{\hat\sL_+}, g_{\hat\sL_+})_{\cC}-(f_{\hat\sL_-}, g_{\hat\sL_-})_{\cC}, \qquad f=f_{\hat\sL_+}+f_{\hat\sL_-}, \quad g=g_{\hat\sL_+}+g_{\hat\sL_-} $$ which coincides with the original indefinite inner product $[\cdot,\cdot]$ on $\cD$.

\section{$J$-self-adjoint operators with property of $\cC$-symmetry}
 \begin{definition}\label{ww7b}
 A densely defined operator $H$ in $\sH$
has \emph{the property of ${\cC}$-symmetry}
if there exists an operator ${\cC}$ with properties described in Theorem \ref{ww1b} and such that
\begin{equation}\label{new6b}
H{{\cC}}f={\cC}Hf, \qquad \forall{f}\in{D}(H).
\end{equation}
\end{definition}

We will say that $H$ has \emph{bounded (unbounded)} $\cC$-symmetry if the corresponding operator $\cC$ is bounded (unbounded).

The commutation identity (\ref{new6b}) requires an additional explanation because $\cC$ may be unbounded.
Precisely, if (\ref{new6b}) is satisfied, then
$$
D(\cC)\supset{D}(H), \quad \cC : D(H)\to{D}(H), \quad  H: D(H)\to{D}(\cC).
$$
If $\cC$ is a bounded operator, then the first and third relations are trivial since $D(\cC)=\sH$.

The existence of $\cC$-symmetry for $H$ means that the $J$-orthogonal direct sum (\ref{e8}), where $\sL_\pm=\frac{1}{2}(I\pm\cC)\sH$
decomposes the operator $H$ into the sum
\begin{equation}\label{new25}
H=H_+\dot{+}{H_-}, \quad D(H)=D(H_+)\dot{+}D(H_-),  \quad D(H_{\pm})\subset\sL_\pm
\end{equation}
of operators $H_{\pm} : D(H_{\pm})\to\sL_\pm$ acting in $\sL_\pm$.

If $H$ is a $J$-self-adjoint ($J$-symmetric), then its components $H_{\pm}$ in (\ref{new25})
should be $J$-self-adjoint ($J$-symmetric) operators in $\sL_{\pm}$. In this case
$H$ turns out to be a self-adjoint (symmetric) operator with respect to the new inner product $(\cdot,\cdot)_{\cC}$.
Therefore,  \emph{the description of $\cC$-symmetry for a $J$-self-adjoint ($J$-symmetric) operator $H$ leads to the explicit construction of the new inner product $(\cdot,\cdot)_{\cC}$ which ensures the self-adjointness (the symmetricity) of $H$}.

\begin{proposition}\label{p1}
Let $H$ be a $J$-self-adjoint operator with non-empty resolvent set $\rho(H)\not=\emptyset$. Then $H$ may only have a property of \emph{bounded} $\cC$-symmetry.
\end{proposition}
\emph{Proof.} If $H$ is $J$-self-adjoint, then $H$ should be a closed operator in $\sH$. Let $\lambda\in\rho(H)$. Then
the operator $H-\lambda{I}$ commutes with the same operator $\cC$ and ${D(\cC)}\supseteq\cR(H-\lambda)=\sH$.
Therefore, $\cC$ is a bounded operator. \rule{2mm}{2mm}

\begin{remark}{}\

{1.}  Proposition \ref{p1} may seem paradoxical only at the first glance. Indeed, if $\cC$ is unbounded, then the new inner product $(\cdot,\cdot)_{\cC}$
is singular with respect to the original inner product $(\cdot,\cdot)$. Therefore, the closedness of $H$  with respect to $(\cdot,\cdot)$
together with the condition $\rho(H)\not=\emptyset$ do not fit the singular inner product $(\cdot,\cdot)_{\cC}$. For this reason, the property of unbounded $\cC$-symmetry is more natural for the case of closable (but no closed) operators.

{2.}  The condition $\rho(H)\not=\emptyset$ is important in Proposition \ref{p1}. Indeed, every unbounded operator $H=\cC$ is
closed (since $\cC$ is $J$-self-adjoint), its spectrum coincides with $\mathbb{C}$ and, evidently, this operator  has the property of unbounded $\cC$-symmetry.
\end{remark}

The next result is a direct consequence of Theorem 1.9 in \cite{AK_AK} and Proposition \ref{p1}.

\begin{theorem}\label{ww2}
Let $H$ be a $J$-self-adjoint operator. Then the following assertions are equivalent:
\begin{itemize}
  \item[(i)] $H$ has the property of $\cC$-symmetry;
 \item[(ii)] $H$ is a self-adjoint operator with respect to a new (equivalent to $(\cdot,\cdot)$) inner product in the Hilbert space
$(\sH, (\cdot,\cdot))$;
  \item[(iii)] $H$ is similar\footnote{an operator $H$ is called similar to
a self-adjoint operator $A$ if there exists a bounded and boundedly invertible operator
$Z$ such that $ZA = HZ$.} to a self-adjoint operator in $(\sH, (\cdot,\cdot))$.
\end{itemize}
\end{theorem}

The item $(iii)$ of Theorem \ref{ww2} allows one to
characterize the (possible) existence of $\cC$-symmetry for a $J$-self-adjoint operator $H$  with the
use of a well-known criterion of similarity (see, e.g., \cite{Naboko}).
Precisely, a $J$-self-adjoint operator $H$ has the property of $\cC$-symmetry
if and only if it has a real spectrum  and
there exists a constant $M$ such that
 \begin{equation}\label{bebe84}
 \mathrm{sup}_{\varepsilon>0}\varepsilon\int_{-\infty}^{\infty}\|(H-zI)^{-1}f\|^2d\xi\leq{M}\|f\|^2, \quad  \forall{f}\in\mathfrak{H},
 \end{equation}
 where the integral is taken along the line $z=\xi+i\varepsilon$ ($\varepsilon>0$
 is fixed).

However, the relation (\ref{bebe84}) does not answer how to construct the corresponding operator $\cC$.

The explicit formulas for $\cC$ can be derived for special classes of $J$-self-adjoint operators:

{\bf I.} Let us assume that a $J$-self-adjoint operator $H$ is $J$-nonnegative, i.e., $[Hf,f]\geq{0}$ for all $f\in{D}(H)$.
If the resolvent set $\rho(H)$  is nonempty, then the spectrum of $H$ is real \cite[Theorem 3.27]{AK_Azizov}.
Moreover, $H$ has a spectral function $E_H(\cdot)$ with
properties similar to that of a spectral function of a self-adjoint operator  \cite{L3}.
The main difference is the occurrence of critical points $\{0, \infty\}$. Note that
$\infty$ is always a critical point of $H$, and $0$ may be its critical point.
Significantly different behavior of the spectral function $E_H(\cdot)$
occurs at \emph{singular} critical points in any neighborhood of which $E_H(\cdot)$ is unbounded.

The next statement is a direct consequence of \cite[Theorem 5.7]{L3}

\begin{proposition}\label{wewe}
Let $H$ be a $J$ -nonnegative and $J$ -self-adjoint operator. Assume that the resolvent set $\rho(H)$ is non-empty,
$0$ and $\infty$ are not singular critical points and $0$ is not an eigenvalue of $H$. Then
$H$ has a bounded $\cC$-symmetry $\cC=E_H(\mathbb{R}_+)-E_H(\mathbb{R}_-)$, where $E_H(\cdot)$ is the
spectral function of $H$.
\end{proposition}

The operator
$$
H=(\mbox{sgn} \ x)\frac{d^2}{dx^2}
$$
defined on its maximal domain in $L_2(\mathbb{R})$ is $J$ -nonnegative and $J$ -self-adjoint with
the fundamental symmetry $Jf=(\mbox{sgn} \ x)f$ in $L_2(\mathbb{R})$. It was showed \cite{Branko}
that $H$ satisfies all assumptions of Proposition \ref{wewe}. Hence, $H$ has a bounded $\cC$-symmetry.

{\bf II.} Let $H$ be a $J$-self-adjoint operator with Riesz basis of eigenvectors $\{f_n\}$ corresponding to
real simple eigenvalues. Then \cite[Theorem 6.3.7]{AK_AK}, $H$ has the property of $\cC$-symmetry,
the corresponding bounded operator $\cC$ is determined uniquely and
$$
\cC{f}=\sum_{n=1}^\infty{[f,f_n]f_n}, \qquad f\in\sH.
$$
(Without loss of generality, in the last formula,  we assume that $\{f_n\}$  are
\emph{normalized with respect to the indefinite metric} $[\cdot,\cdot]$, i.e.,
$[f_n,f_n]^2=1$.)

\section{$\cC$-symmetry for $J$-symmetric operators with a complete set of eigenvectors}

\subsection{Description of operators $\cC$ associated with complete set of eigenvectors.}
Let a $J$-symmetric operator $H$ have a complete set of eigenvectors $\{f_n\}_{n=1}^{\infty}$ corresponding to
real simple eigenvalues $\{\lambda_n\}_{n=1}^{\infty}$. It should be noted that $[f_n, f_n]\not=0$ for any eigenfunction
$f_n$. This fact is crucial for our investigations. Let us verify it. Indeed, it is well known that eigenfunctions corresponding to
different real eigenvalues of a $J$-symmetric operator are $J$-orthogonal \cite{AK_Azizov}. Hence, if there exists an eigenfunction
$f_m$  such that $[f_m, f_m]=0$, then $f_m$ will be $J$-orthogonal to the linear span $\mathcal{S}$ of $\{f_n\}$.
The latter means that $Jf_m$ is orthogonal to the dense set $\mathcal{S}$ in the Hilbert space $(\sH, (\cdot,\cdot)$.
Therefore, $Jf_m=0$ and $f_m=0$ that is impossible.

Separating the sequence of eigenvectors $\{f_n\}$ by the signs of $[f_n,f_n]$:
\begin{equation}\label{bebe95}
f_{n}=\left\{\begin{array}{l}
f_{n}^+ \quad \mbox{if} \quad [f_{n},f_{n}]>0, \\
f_{n}^- \quad \mbox{if} \quad [f_{n},f_{n}]<0
\end{array}\right.
\end{equation}
we obtain two sequences of positive $\{f_n^+\}$ and negative $\{f_n^-\}$ elements of the Krein space $(\mathfrak{H}, [\cdot,\cdot])$.

Let ${\mathfrak L}_+^0$ and ${\mathfrak L}_-^0$ be the closure (with respect to the initial inner product $(\cdot,\cdot)$) of
the linear spans generated by the sets $\{f_n^+\}$ and $\{f_n^-\}$, respectively. By construction, ${\mathfrak L}_\pm^0$ are $J$-orthogonal positive/negative subspaces in $(\mathfrak{H}, [\cdot,\cdot])$ and the direct sum
\begin{equation}\label{ww15}
{\mathcal D}_0={\mathfrak L}_+^0[\dot{+}]{\mathfrak L}_-^0
\end{equation}
 is a dense set in ${\mathfrak H}$ (since ${\mathcal D}_0$ contains ${\mathcal S}$).

Let $\cC_0$ be an operator associated with (\ref{ww15}):
\begin{equation}\label{AK43}
 \cC_0f=f_{{\mathfrak L}_{+}^0}-f_{{\mathfrak L}_{-}^0}, \qquad f\in{D}(\cC_0)=\mathcal{D}_0.
\end{equation}

It follows from (\ref{AK43}) that the operator $\cC_0$ is densely defined and $\cC_0^2f=f$ for any $f\in\mathcal{D}(\cC_0)$.
However, in general case, the operator $\cC_0$ \emph{cannot be considered} as an operator of $\cC$-symmetry.
The matter is that the subspaces ${\mathfrak L}_\pm^0$ in (\ref{ww15}) may be {\it proper subspaces of a maximal definite subspaces} ${\mathfrak L}_\pm$.

If at least one of subspaces ${\mathfrak L}_\pm^0$ loses the property of being maximal in the classes of definite subspaces (positive or negative), then the sum (\ref{ww15}) \emph{cannot be domain of definition} of an operator of $\cC$-symmetry with properties described in Theorem \ref{ww1b}. In other words, the operator $\cC_0$ cannot be presented as $Je^Q$, where $e^Q$ is a \emph{self-adjoint operator}.
Of course, this phenomenon is possible only in the case of unbounded operators $\cC$.

In order to construct a proper operator $\cC$ associated with (\ref{ww15})
we have to extend (\ref{ww15}) to a $J$-orthogonal direct sum ${\mathfrak L}_+[\dot{+}]{\mathfrak L}_-$,
where ${\mathfrak L}_\pm\supset{\mathfrak L}_\pm^0$ are \emph{maximal} definite subspaces (positive and negative).
Obviously, the operator $\cC$ determined by the direct sum ${\mathfrak L}_+[\dot{+}]{\mathfrak L}_-$ will be an extension of $\cC_0$.

In general, the mentioned extension
\begin{equation}\label{bebe85}
{\mathfrak L}_+^0[\dot{+}]{\mathfrak L}_-^0  \ \rightarrow \ {\mathfrak L}_+[\dot{+}]{\mathfrak L}_-
\end{equation}
is not determined uniquely\footnote{this phenomenon was discovered by Langer \cite{R12}}.
This leads to the \emph{nonuniqueness} of unbounded operators of $\cC$-symmetry $\cC\supset\cC_0$ associated with ${\mathfrak L}_+^0[\dot{+}]{\mathfrak L}_-^0$. Therefore, the complete set of eigenvectors $\{f_n\}$ does not always determine a unique
operator $\cC$.

In order to describe all possible extensions (\ref{bebe85}) we introduce the operator
$G_0=J\cC_0$, where $\cC_0$ is defined by (\ref{AK43}).

\begin{lemma}
The operator $G_0$ is a closed positive symmetric operator in $(\sH, (\cdot,\cdot))$ and
\begin{equation}\label{AK71}
JG_0JG_0f=f, \qquad \forall{f}\in{\mathcal D}_0={D}(G_0)={D}(\cC_0).
\end{equation}
\end{lemma}
\emph{Proof.}
Repeating the proof of Lemma 6.2.3 in \cite{AK_AK} we decide that
$\cC_0$ is a closed operator in $\sH$. Therefore, the operator $G_0=J\cC_0$ is closed too.
The positivity of $G_0$ follows from the fact that
$$
(G_0f,f)=[\cC_0{f},f]=[f_{{\mathfrak L}_{+}^0}, f_{{\mathfrak L}_{+}^0}]-[f_{{\mathfrak L}_{-}^0}, f_{{\mathfrak L}_{-}^0}]>0
$$
for all nonzero $f\in{\mathcal D}_0$. The `boundary condition'
(\ref{AK71}) is the direct consequence of the relation
 $\cC_0^2=I$ on $D(\cC_0)$. \rule{2mm}{2mm}

The next result reduces the description of all possible operators $\cC$ associated with the initial direct sum
(\ref{ww15}) to the description of a special class of positive self-adjoint extensions $G$ of $G_0$.

\begin{proposition}\label{new56}
The set of all positive self-adjoint extensions $G$ of $G_0$  satisfying
the condition
\begin{equation}\label{AK72}
JGJGf=f, \qquad \forall{f}\in{D}(G)
\end{equation}
is in one-to-one correspondence with the set of all possible extensions
(\ref{bebe85}) where $\sL_\pm$ are $J$-orthogonal maximal positive/negative
subspaces in the Krein space $(\sH, [\cdot,\cdot])$.
\end{proposition}
\emph{Proof.} Let $G$ be a positive self-adjoint extension of $G_0$ such that (\ref{AK72}) holds.
Then $\cC=JG$  satisfies assertion (iii) in Theorem \ref{ww1b}.
Therefore, the operator $\cC$ determines the $J$-orthogonal maximal positive/negative subspaces $\sL_\pm$
by the formula (\ref{e9}). These subspaces are extensions of ${{\mathfrak L}_{\pm}^0}$
since $\cC$ is an extension of $\cC_0$.

Conversely, each extension (\ref{bebe85}) defines an operator $\cC$ by formula
(\ref{new5}). This operator is an extension of $\cC_0$ and, hence the positive self-adjoint operator $G=J\cC$ is an extension of $G_0=J\cC_0$. \rule{2mm}{2mm}

\subsection{The case of Friedrichs extension.}
Let ${D}[G_0]\subset\mathfrak{H}$ be the closure of
${D}(G_0)$ with respect to the norm
\begin{equation}\label{ww31}
\|f\|^2_0=\|f\|^2+(G_0f,f), \qquad  f\in{D}(G_0).
\end{equation}

The operator $G_F$ defined as follows:
$$
G_Ff=G_0^*f, \qquad f\in{D}(G_F)={D}[G_0]\cap{D}(G_0^*)
$$
is called the Friedrichs extension of $G_0$.

The Friedrichs extension $G_F$ preserves the lower bound of $G_0$:
$$
\displaystyle{\inf_{f\in{D}(G_0)\setminus\{0\}}\frac{(G_0f,f)}{(f,f)}=\inf_{f\in{D}(G_F)\setminus\{0\}}\frac{(G_F{f},f)}{(f,f)}}.
$$
and it is a unique positive self-adjoint extension having the domain in $D_0[G_0]$ \cite{ArTs, AK_Krein}.

\begin{theorem}\label{new67}
The Friedrichs extension $G_F$ satisfies the relation
$$
JG_FJG_Ff=f, \qquad f\in{D}(G_F)
$$
if and only if $G_F$ is the unique positive self-adjoint extension of $G_0$.
\end{theorem}
\emph{Proof.} The operator $G_0$ has the inverse $G_0^{-1}$  which is defined on the dense set $JD(G_0)$
in $\sH$ and the condition (\ref{AK71}) can be rewritten as
\begin{equation}\label{ww36}
JG_0f=G_0^{-1}Jf,  \qquad f\in{D}(G_0).
\end{equation}

By virtue of (\ref{ww36}) the operator $J$ isometrically maps the Hilbert
space $(D[G_0], \|\cdot\|_0)$ onto the Hilbert space $(D[G_0^{-1}], \|\cdot\|_1)$, where
$D[G_0^{-1}]$ is the completion of the domain of definition $D(G_0^{-1})=JD(G_0)$ with respect to the norm
(cf (\ref{ww31}))
$$
\|g\|^2_1=\|g\|^2+(G_0^{-1}g,g), \qquad g{\in}JD(G_0).
$$

The calculation of adjoint operators in (\ref{ww36}) gives
$G_0^*J=J(G_0^{-1})^*$ or $JG_0^*=(G_0^{-1})^*J$. Therefore, for any $f\in{D(G_F)}=D[G_0]\cap{D(G_0^*)}$,
$$
JG_0^*f=JG_Ff=(G_0^{-1})^*Jf=(G_0^{-1})_FJf,
$$
where $Jf\in{D[G_0^{-1}]}\cap{D((G_0^{-1})^*)}$ and $(G_0^{-1})_F$ is the Friedrichs extension of the positive symmetric
operator $G_0^{-1}$. Using the well-known relation $(G_0^{-1})_F=(G_K)^{-1}$ (see, e.g. \cite{ArTs}), where
$G_K$ is the Krein-von Neumann extension of $G_0$,
we conclude that
$$
JG_Ff=(G_K)^{-1}Jf, \qquad f\in{D}(G_F).
$$
Therefore, the required condition $JG_FJG_F=I$ holds if and only if the Friedrichs extension $G_F$ of $G_0$
coincides with the Krein-von Neumann extension $G_K$. This is possible if and only if $G_0$ has a unique
positive self-adjoint extension \cite{AK_Krein}.
\rule{2mm}{2mm}

\begin{corollary}
The operator $\cC=JG_F$ is the operator of $\cC$-symmetry associated with (\ref{ww15})
if and only if
\begin{equation}\label{bebe87}
\inf_{f\in{D}(G_0)\setminus\{0\}}\frac{(G_0f,f)}{|(f,g)|^2}=0
\end{equation}
for all nonzero vectors $g\in\ker(I+G_0^*)$.
\end{corollary}

\emph{Proof.} It follows from Proposition \ref{new56} and
Theorem \ref{new67} that the operator $\cC=JG_F$  determines a $J$-orthogonal sum of maximal definite
subspaces $\mathfrak{L}_{\pm}$ which is the extension of (\ref{ww15}) if and only if
$G_F$ is the unique positive self-adjoint extension of $G_0$.  The latter is equivalent
to the condition (\ref{bebe87}) \cite[Theorem 9]{AK_Krein}.
\rule{2mm}{2mm}

Let $\cC$ be an operator of $\cC$-symmetry associated with (\ref{ww15})
and let $\mathfrak{H}_{\cC}$ be the corresponding Hilbert space obtained by the completion of
${\mathfrak L}_+[\dot{+}]{\mathfrak L}_-$  with respect to the inner product $(\cdot, \cdot)_{\cC}$
(see (\ref{e8c})).

By construction, the original direct sum ${\mathfrak L}_+^0[\dot{+}]{\mathfrak L}_-^0$
determined by the eigenfunctions $\{f_n\}$ is a linear manifold in the new Hilbert space $(\mathfrak{H}_{\cC}, (\cdot, \cdot)_{\cC})$.
However, in  the general case, we cannot state that this linear manifold is a dense set in $\mathfrak{H}_{\cC}$.
\begin{theorem}\label{bebe90}
If the operator $\cC$ is determined by the Friedrichs extension $G_F$ (i.e., if the condition (\ref{bebe87}) holds),
then the linear span of eigenvectors $\{f_n\}$ is a dense set in $\mathfrak{H}_{\cC}$
\end{theorem}
\emph{Proof.}
The Friedrichs extension $G_F$ is an example of \emph{extremal extension} of $G_0$ \cite{AK_Arlin}.
By definition, a nonnegative self-adjoint extension  $G$ of
$G_0$ belongs to the class $E(G_0)$ of extremal extensions if
\begin{equation}\label{bebe86}
\inf_{f\in{D}(G_0)}{(G(\phi-f),(\phi-f))}=0, \quad \mbox{for all} \quad \phi\in{D}(G).
\end{equation}

Since $\|\phi-f\|^2_{\cC}=[\cC(\phi-f), \phi-f]=(G(\phi-f),(\phi-f))$, the condition (\ref{bebe86})
means that each element $\phi\in{D}(\cC)={D}(G)$ can be approximated by elements
$f\in{D}(G_0)={\mathfrak L}_+^0[\dot{+}]{\mathfrak L}_-^0$ in $\sH_{\cC}$.
Now, in order to complete the proof, it suffices to verify that the span of $\{f_n\}$ is a dense subset
of ${D}(G_0)$ with respect to $(\cdot,\cdot)_{\cC}$.

The inner product $(\cdot,\cdot)_{\cC}$ considered on ${\mathfrak L}_\pm^0$  coincides with $\pm[\cdot,\cdot]$ and satisfies the evaluation
$$
(f_{{\mathfrak L}_\pm^0}, f_{{\mathfrak L}_\pm^0})_{\cC}=|(\cP{f}_{{\mathfrak L}_\pm^0}, f_{{\mathfrak L}_\pm^0})|\leq\|\cP{f}_{{\mathfrak L}_\pm^0}\|\|f_{{\mathfrak L}_\pm^0}\|=\|f_{{\mathfrak L}_\pm^0}\|^2=(f_{{\mathfrak L}_\pm^0}, f_{{\mathfrak L}_\pm^0})
$$
for any $f_{{\mathfrak L}_\pm^0}\in{\mathfrak L}_\pm^0$.
Therefore, each element $f_{{\mathfrak L}_\pm^0}\in{{\mathfrak L}_\pm^0}$ can be approximated by
$\mbox{span}\{f_n^\pm\}$ with respect to $(\cdot,\cdot)_{\cC}$  since
${{\mathfrak L}_\pm^0}$ are closure of $\mbox{span}\{f_n^\pm\}$ with respect to $(\cdot,\cdot)$.
This means that $\mbox{span}\{f_n\}=\mbox{span}\{f_n^+\}\cup\mbox{span}\{f_n^-\}$
is a dense set in $D(G_0)$ with respect to $(\cdot,\cdot)_{\cC}$. \rule{2mm}{2mm}

\begin{corollary}
If the operator $\cC$ is determined by the Friedrichs extension $G_F$,
then the normalized eigenvectors\footnote{i.e., $\{f_n\}$ are normalized with respect to the indefinite metric $[\cdot,\cdot]$:
$[f_n,f_n]^2=1$} $\{f_n\}$ forms an orthonormal basis of the Hilbert space  $(\mathfrak{H}_{\cC}, (\cdot, \cdot)_{\cC})$.
\end{corollary}
\emph{Proof.} By virtue of (\ref{new7}) and (\ref{AK43}), the normalized eigenvectors
$\{f_n\}$ are an orthonormal system in the Hilbert space $(\mathfrak{H}_{\cC}, (\cdot, \cdot)_{\cC})$.
Due to Threorem \ref{bebe90}, the linear span of $\{f_n\}$ is a dense set in $\mathfrak{H}_{\cC}$. Therefore,
$\{f_n\}$ turns out to be an orthonormal basis of this Hilbert space. \rule{2mm}{2mm}

Up to now we do not specify the domain of definition of the $J$-symmetric operator $H$.
Let the domain $D(H)$ coincide with the linear span of $\{f_n\}$. Then relation (\ref{new6b})
holds for any operator $\cC$ associated with (\ref{ww15}). If the Friedrichs extension $G_F$
determines an operator of $\cC$-symmetry (i.e., if condition (\ref{bebe87}) holds), then
the operator $H$ turns out to be essentially self-adjoint in the Hilbert space $(\sH_{\cC_F}, (\cdot,\cdot)_{\cC_F})$
where $\cC_F=JG_F$ and $\sigma(H)$ coincides with the closure of $\{\lambda_n\}$.

\begin{example}\emph{Schauder basis of eigenvectors.}

Let normalized eigenvectors $\{f_n\}_{n=1}^{\infty}$ of a $J$-symmetric operator $H$
with real simple eigenvalues form a Schauder basis.
We recall that a sequence $\{f_n\}$ is called \emph{a Schauder basis} of a Hilbert space
$\sH$ if, for each $f\in\sH$, there exist uniquely determined scalar coefficients $\{c_n \}$ such that
$$
f=\sum_{n=1}^\infty{c_nf_n}.
$$

The coefficients $\{c_n \}$ can easy be specified with the use
of a biorthogonal sequence\footnote{biorthonormality of seguences $\{f_n\}$ and $\{g_n\}$ means that $(f_n, g_m)=\delta_{nm}$}  $\{g_n\}$,
namely: $c_n=(f, g_n)$. Since the eigenvectors $\{f_n\}$ are mutually $J$-orthogonal and normalized,
the sequence $\{g_n\}$ has the form $g_n=[f_n,f_n]Jf_n$. Therefore, taking into account
the separation (\ref{bebe95}) of eigenvectors $\{f_n\}$ by the signs of $[f_n,f_n]$, we obtain
$$
f=\sum_{n=1}^\infty[f_n, f_n][f, f_n]{f_n}=\sum_{n=1}^\infty[f, f_n^+]{f_n^+}-\sum_{n=1}^\infty[f, f_n^-]{f_n^-}, \quad \forall{f}\in\sH.
$$

If $\{f_n\}$ is a Schauder basis, then the subspaces ${\mathfrak L}_\pm^0$
defined as the closure of the linear spans of $\{f_n^\pm\}$, respectively
are maximal positive/negative in the Krein space $(\sH, [\cdot,\cdot])$ \cite[Statement 10.12 in Chapter 1]{AK_Azizov}.
Therefore, the formula (\ref{AK43}) determines an unbounded operator $\cC=\cC_0$
with the domain of definition (\ref{ww15}). For each $f\in{D}(\cC)=\mathcal{D}_0$,
$$
\cC{f}=f_{\sL_+^0}-f_{\sL_-^0}=\sum_{n=1}^\infty[f_{\sL_+^0}, f_n^+]{f_n^+}+\sum_{n=1}^\infty[f_{\sL_-^0}, f_n^-]{f_n^-}=\sum_{n=1}^\infty[f, f_n]{f_n}.
$$

The operator $G_0=J\cC$ is a positive self-adjoint operator. Therefore, its Friedrichs extension coincides with $G_0$.

If the domain $D(H)$ coincides with the linear span of $\{f_n\}$, then $H$ has the property of $\cC$-symmetry
with the operator $\cC$ determined above. The operator $H$ will be essentially self-adjoint in the new Hilbert space $(\sH_{\cC}, (\cdot,\cdot)_{\cC})$.
\end{example}

 \end{document}